# Tunable Asymmetric Acoustic Absorption in Ventilated Metasurfaces

Keqiang Lyu[1], Mohamed Farhat[2], and Ying Wu[1,2]*

[1]Physical Science and Engineering (PSE) Division, King Abdullah University of Science and Technology (KAUST), Thuwal 23955-6900, Saudi Arabia

[2]Computer, Electrical and Mathematical Science and Engineering (CEMSE) Division, King Abdullah University of Science and Technology (KAUST), Thuwal 23955-6900, Saudi Arabia

*Contact author: ying.wu@kaust.edu.sa

Asymmetric sound absorption is essential for advanced acoustic manipulation. However, current frequency modulation and broadbanding highly depend on geometric reconfiguration, leading to inevitable structural complexity that impedes their practical applications. Here, we propose a tunable, highly efficient, asymmetric ventilated acoustic system comprising two heterogeneous resonators. Specifically, it couples a highly dissipative space-coiling resonator (SCR) as a dark mode for energy consumption, alongside a weakly damped Helmholtz resonator as a bright mode acting as a reflective soft boundary. Theoretical and numerical analyses reveal strong asymmetry within the deep-subwavelength region (with a resonator size of approximately $\lambda/9.4$), achieving 99% absorption for left-incident waves and 98% reflection for right-incident ones. Furthermore, the SCR introduces an interesting degree of freedom for acoustic tuning. Simply rotating the resonator induces a 92% absorption drop (~11 dB attenuation), functioning as an "Acoustic Switch". Moreover, this rotation significantly shifts the operating band. By parallel-coupling multi-angle isomorphic resonators, we achieve efficient broadband absorption (>0.8) from 325 to 375 Hz, offering an attractive paradigm for tunable acoustic metasurfaces and ventilated absorbers.

## I. INTRODUCTION

Sound absorption plays a crucial role in mitigating environmental noise and is widely applied in fields such as architecture, machinery, and pipeline systems. Traditional sound absorption methods typically rely on the combination of rigid backing structures and porous materials, which can effectively reduce noise in the mid-to-high frequency range, but their absorption performance in the low-frequency regime remains insufficient due to the mass-density law [1]. Acoustic metamaterials can manipulate sound waves in unconventional ways, revolutionizing acoustic design and engineering by significantly enhancing the control over sound propagation [2]. Compared to traditional absorbers, these structures and materials exhibit superior performance. These sound-absorbing structures are primarily composed of subwavelength resonant units, such as membrane resonators [3–7], coiling-up space structures [8–13], and helmholtz resonators [14–17], which can efficiently absorb and dissipate acoustic energy. However, the rapid development of modern acoustics has imposed more stringent practical requirements on sound-absorbing structures. Traditional single-port designs have inevitably impeded the transfer of airflow and heat, thereby shifting research attention toward two-port systems [18,19] with free air circulation capabilities. Furthermore, in certain specific scenarios, the system not only needs to maintain ventilation and suppress external incoming noise but also generate specific reflections for internal sound waves. Therefore, developing devices with both ventilation and asymmetric sound absorption capabilities has become a major hotspot in the current field of acoustic metamaterials.

To achieve this goal of asymmetric sound absorption, researchers have proposed various innovative solutions based on metamaterials. Typical designs include asymmetric channels integrating lossy structures, folded Fabry-Perot resonators, and narrow slit channels [20], Helmholtz resonator pairs based on bright and dark mode coupling [15,21–26], detuned resonators embedded in waveguides [27], periodic structures filled with foam to reconstruct surface impedance [28], coiling structure resonators combining high dissipation and strong radiation loss [29], and hybrid membrane resonators [4] with lossy Bragg stacks [30]. Furthermore, by utilizing exceptional points [31] in pt-symmetric systems and bound states in the continuum (BIC) [32] to construct acoustic non-Hermitian systems, unidirectional perfect acoustic absorption has also been successfully achieved. Although these designs have achieved remarkable success, they still face certain limitations in practical applications. For instance, some systems suffer from relatively high working frequencies [23,27,29,32] or small ventilation areas [27,33,34], and the membrane surface tension [4] or intrinsic loss factors [25,35] are often difficult to control accurately. A more severe challenge lies in the fact that the working frequencies of almost all existing designs are fixed. Once it is necessary to change the target absorption frequency or achieve broadband absorption, the geometric dimensions of the resonators must usually be altered (such as cascading an array of resonators with gradient dimensions [20,22,24,27,34,35]). This design paradigm, which relies heavily on adjusting internal structural parameters, greatly increases the complexity of engineering manufacturing and processing costs.

Breaking away from traditional paradigms, this study constructs an asymmetric acoustic absorption ventilation system by coupling a highly dissipative space-coiling resonator (SCR) (acting as a dark mode to dissipate acoustic energy) with a weakly damped classical Helmholtz resonator (acting as a bright mode to provide soft-boundary reflection) within a two-port waveguide. Based on the temporal coupled-mode theory (TCMT), we precisely tailored the dimensions of each resonator to achieve co-frequency resonance and bright-dark mode coupling, thereby perfectly realizing the asymmetric absorption and reflection of acoustic waves within this two-port architecture. Theoretical and numerical analyses reveal profound asymmetry at 365 Hz (size $\lambda/9.4$), achieving 99% left-incident absorption and 98% right-incident reflection. Crucially, the SCR introduces an interesting tuning degree of freedom: simply rotating it without changing internal parameters drops absorption at 365 Hz by 92% (nearly 11 dB attenuation), offering a viable approach for continuously tunable devices. Furthermore, by parallel-coupling multiple SCRs with distinct resonant frequencies induced by varying rotation angles, we expanded the system's performance to achieve broadband absorption (>0.8) from 325 Hz to 375 Hz. This characteristic of "identical structure, multiple responses" simplifies the design of broadband acoustic metasurfaces from "customization of structural morphology" to "combination of rotational parameters." This greatly lowers the manufacturing threshold and provides a completely useful method for the standardized mass production and reconfigurable design of acoustic devices.

## II. RESULTS AND DISCUSSION

**2.1 Theoretical framework and design concepts**

As schematically illustrated in Fig. 1(a), we propose an asymmetric acoustic absorber composed of two distinct resonators. By sharing the same resonance frequency ($\omega_1 = \omega_2$) but possessing different leakage and loss factors, the system is designed for the unidirectional routing of acoustic energy. Specifically, it yields near-perfect absorption for left incidence and near-perfect reflection for right incidence. The concepts of bright and dark modes originated in quantum dynamics to characterize the absorption and emission properties of atomic or molecular systems interacting with photons [36]. This theoretical framework was subsequently adapted into electromagnetics and plasmonics to evaluate the coupling efficiency between incident electromagnetic waves and localized modes [37,38]. Inspired by this, we define an acoustic bright mode as a resonator with high leakage and low loss, and a dark mode as one with low leakage and high loss. This framework provides a new physical route for asymmetric acoustic wave manipulation [26,39]. As depicted in Fig. 1(a), the proposed bright mode(right) functions as an acoustic soft boundary, producing a nearly out-of-phase reflection, while the dark mode (left) absorbs the bulk of the incident energy due to its high loss factor.

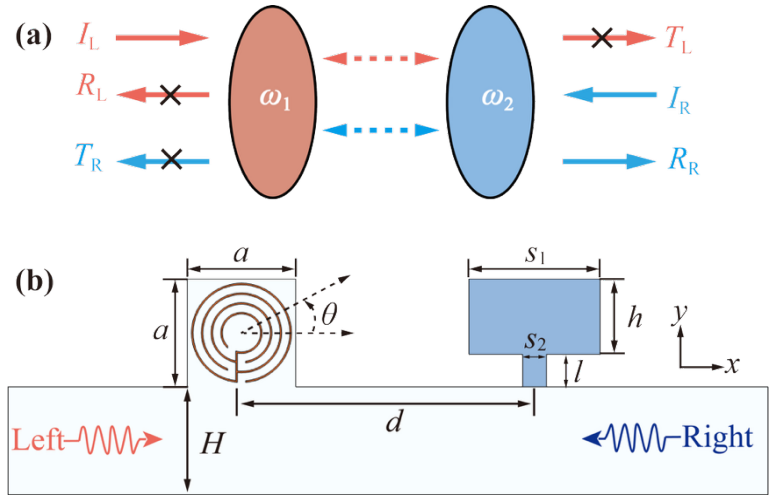

**Fig. 1. Low-frequency asymmetric sound absorber.** (a) Schematic of asymmetric absorber constructed by coupling dark-bright modes with identical resonant frequencies. (b) Designed asymmetric absorber with geometric parameters annotated.

Before investigating the overall absorption performance of the system depicted in Fig. 1(a), it is essential to independently analyze a single waveguide-coupled mode. The total loss of this mode is dictated by its resonance half-width $\delta\omega_r$, and resonance frequency $\omega_r$, formulated as [40] $Q^{-1} = \delta\omega_r/\omega_r = Q_{\text{leak}}^{-1} + Q_{\text{loss}}^{-1}$, where $Q_{\text{leak}}^{-1}$ and $Q_{\text{loss}}^{-1}$ represent the leakage and intrinsic loss factors, respectively. By employing the TCMT to characterize the acoustic scattering properties of this two-port system (see Appendix A for details), the complex acoustic pressure reflection and transmission coefficients for the open two-port waveguide loaded with the resonator can be analytically derived as follows [41,42]

$$r = -\frac{Q_{\text{leak}}^{-1}}{2i(\omega/\omega_r - 1) + Q_{\text{leak}}^{-1} + Q_{\text{loss}}^{-1}}, \tag{1}$$

$$t = 1 - \frac{Q_{\text{leak}}^{-1}}{2i(\omega/\omega_r - 1) + Q_{\text{leak}}^{-1} + Q_{\text{loss}}^{-1}}. \tag{2}$$

The energy absorption coefficient is given by $A = 1 - R - T$, where $R = |r|^2$ and $T = |t|^2$ are the energy reflection and transmission coefficient, respectively. According to Eq. (1), at the resonance frequency ($\omega/\omega_r = 1$), the single-resonance system exhibits total reflection provided that $Q_{\text{leak}}^{-1} \gg Q_{\text{loss}}^{-1}$. Furthermore, the combination of Eqs. (1) and (2) dictates that the maximum absorption coefficient [1] of $A_{\max} = 0.5$ is achieved when the leakage and loss factors are perfectly matched ($Q_{\text{leak}}^{-1} \approx Q_{\text{loss}}^{-1}$).

Figure 2(a) shows the variation of the absorption coefficient with $Q_{\text{leak}}^{-1}$ and $Q_{\text{loss}}^{-1}$ at the conventional single-resonance frequency. Calculated using Eqs. (1) and (2), the system achieves near-maximum absorption at the values denoted by the yellow pentagram in Fig. 2(a) ($Q_{\text{leak}}^{-1} = 0.014$, $Q_{\text{loss}}^{-1} = 0.02$), which effectively represents the dark mode for achieving high acoustic absorption. Conversely, at the pentagram in Fig. 2(b) ($Q_{\text{leak}}^{-1} = 0.298$, $Q_{\text{loss}}^{-1} = 0.004$), the power reflection coefficient approaches unity, acting as the bright mode to realize total reflection of acoustic energy.

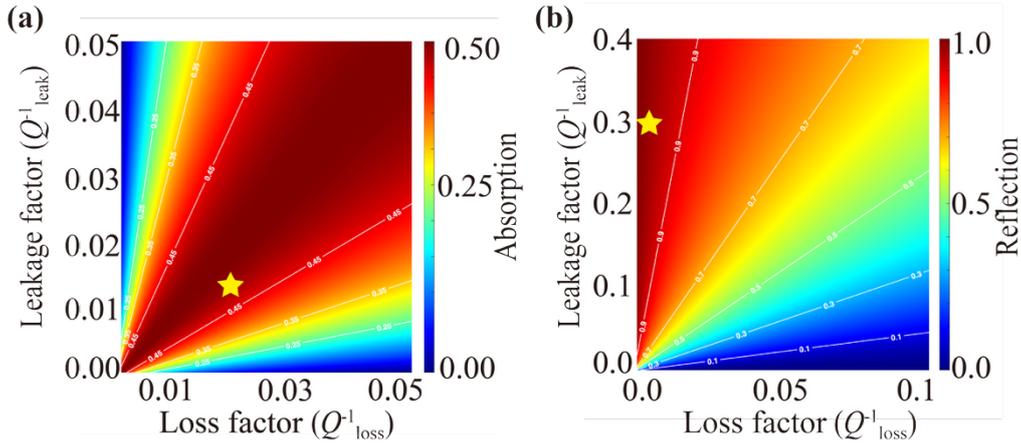

**Fig. 2. Acoustic performance of a single resonator.** (a)The absorption coefficient $A$ and (b) reflection coefficient $R$ modulated by for a single resonator, where the orange pentagrams denote the target conditions at ($Q_{\text{leak}}^{-1} = 0.014$, $Q_{\text{loss}}^{-1} = 0.02$) and ($Q_{\text{leak}}^{-1} = 0.298$, $Q_{\text{loss}}^{-1} = 0.004$), respectively.

### 2.2 Designing acoustic dark and bright mode

Figure 1(b) illustrates the two-dimensional geometric structure of the proposed asymmetric acoustic absorber, tailored to fulfill the target requirements. The absorber consists of a straight waveguide side-coupled with two resonators. The red and blue wavy arrows delineate the propagation paths of acoustic waves incident from the left and right ports, respectively. The left-side unit features a SCR embedded in a square air cavity. The SCR are known to be particularly effective in extending the propagation path within thin structures and enhancing sound absorption at lower frequencies. In fact, the narrowing of the channels intensifies the thermoviscous effects at the air-wall interface, thereby enabling efficient low-frequency absorption with smaller cavity volumes. This composite structure serves as the dark mode in the system, exploiting its inherently high dissipation to efficiently absorb acoustic energy. The structural parameters are set to a cavity side length of $a = 10 \text{ cm}$ and a waveguide height of $H = 10 \text{ cm}$. Notably, this SCR in the dark mode introduces a rotational degree of freedom, $\theta$, defined as the rotation angle of SCR around the center of the square cavity. By tuning $\theta$, the equivalent surface impedance of the acoustic waveguide can be effectively modulated. This mechanism allows for dynamic tuning of both the absorption performance and the operating frequency without requiring any modifications to the internal geometric parameters of the resonator.

The unit positioned on the right side of the waveguide is a classical Helmholtz resonator, operating as the bright mode within the system. It manifests as an equivalent soft boundary with weakly damped characteristics, serving primarily to strongly reflect the incident acoustic energy. The geometric dimensions of this resonator are meticulously configured as follows: neck $s_2 = 2.2 \text{ cm}$, neck length $l=3 \text{ cm}$, cavity width $s_1 = 12.1 \text{cm}$, and cavity height $h = 7 \text{ cm}$. Both resonators are engineered to share an identical resonance frequency, with their center-to-center separation distance set to $d = 27 \text{ cm}$.

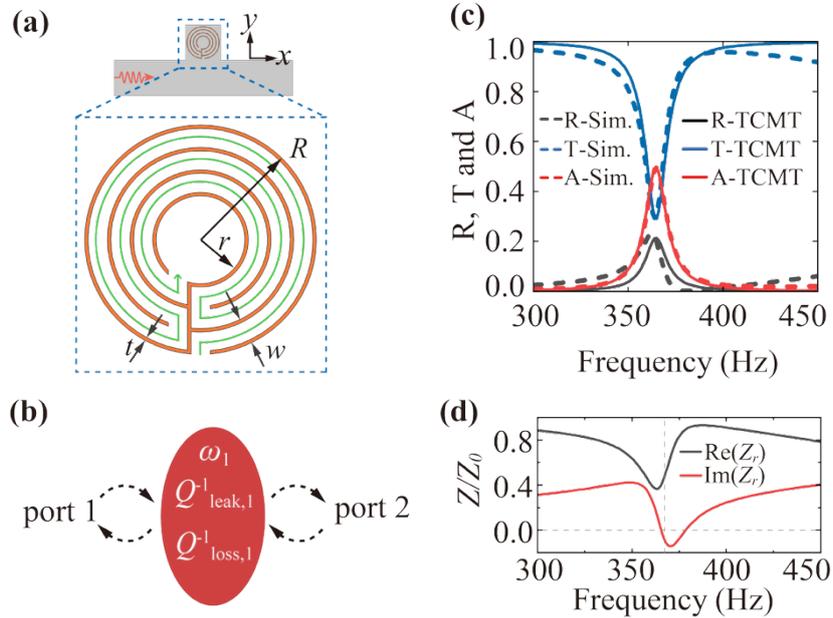

**Fig. 3. The dark mode resonator.** (a)Schematic of the waveguide solely coupled to the dark mode resonator, where the enlarged view reveals the central air cavity and coiled channels of the SCR (b) schematic of the TCMT model. (c) Simulated and theoretical reflection (black curve), transmission (blue curve), and absorption spectra (red curve). (d) Simulated normalized impedance.

As shown in Fig. 3(a), we first analyze the isolated coupling between the main waveguide and the dark-mode resonator. The magnified inset explicitly details the cross-sectional geometry of the SCR, which comprises a central circular air cavity enveloped by coiled channels. Its geometric parameters are configured as follows: outer radius $R = 4.6$ cm, inner cavity radius $r = 1.76$ cm, channel width $w = 0.76$ cm, and solid wall thickness $t = 0.14$ cm. The light gray region represents air, and the yellow region indicates a solid material, such as aluminum or acrylonitrile butadiene styrene plastic—a common three-dimensional (3D) printing material. Notably, the green arrows trace the propagation trajectory of acoustic waves entering the resonator. The folded internal channels substantially elongate the actual propagation path, a geometric feature that is physically equivalent to a reduction in the effective sound speed within the system (i.e., an enhanced equivalent refractive index). Enabled by this space-coiling mechanism, the compact structure operates flawlessly in the low-frequency regime at a deep-subwavelength scale [43]. The mass density and sound speed of air are $\rho_0 = 1.21$ kg/m³ and $c_0 = 343$ m/s, respectively. Figure 3(b) provides a schematic representation of the TCMT framework for the side-loaded system. Governing the energy dynamics, the intrinsic loss of the dark mode is primarily attributed to the severe acoustic dissipation caused by strong thermoviscous effects within the narrow-coiled channels. Meanwhile, the leakage factor dictates the radiative coupling at the interface between the resonator aperture and the main waveguide. To evaluate the acoustic behavior of this dark-mode resonator, we perform finite-element method (FEM) using the frequency-domain solver of COMSOL Multiphysics. The thermoviscous Acoustics module is utilized to account for both thermal and viscous energy dissipation in the narrow channels, in conjunction with the Pressure Acoustics module in the surrounding domains to reduce computational cost. The solid materials, i.e., the yellow areas shown in Fig. 3(a), are assumed to be hard wall because of the large impedance mismatch between air and photosensitive epoxy. The dynamic viscosity of air is $\mu_d = 1.85 \times 10^{-5}$ Pa·s, its thermal conductivity $k = 0.0258$ W/(m·K), and its heat capacity at constant temperature is $C_p = 1005.4$ J/(kg·s).

We first investigate the acoustic properties of the SCR in its initial configuration (i.e., rotation angle $\theta = 0°$). As detailed in Fig. A.1 of Appendix A, the quality factors exhibit distinct dependencies on the structural parameters (channel width $w$ and wall thickness $t$). Specifically, increasing the wall thickness $t$ significantly reduces the intrinsic loss $Q_{loss,1}^{-1}$ while leaving the leakage $Q_{leak,1}^{-1} = 0.014$ nearly unchanged (Fig. A.1(a)). Conversely, widening the channel $w$ leads to a decrease in $Q_{loss,1}^{-1}$ accompanied by a slight increase in $Q_{leak,1}^{-1}$ (Fig. A.1(b)). By comprehensively leveraging these trends and conducting parameter fitting at the target resonance frequency of $f_1 = 365$ Hz, we ultimately extracted and finalized the geometric configuration to achieve the high absorption objective, precisely matching the requisite quality factors: $Q_{leak,1}^{-1} = 0.014$, $Q_{loss,1}^{-1} = 0.02$.

In Fig. 3(c), the solid black, blue, and red curves represent the reflection, transmission, and absorption spectra, respectively, analytically calculated using the TCMT. The corresponding dashed curves denote the FEM results. Around 365 Hz, a prominent absorption peak emerges, accompanied by a drastic drop in transmission and a slight elevation in reflection, indicating partial energy reflection at the resonance frequency. These observations verify that the SCR with a square air cavity effectively functions as a dark mode, significantly enhancing energy dissipation. As derived from Eqs. (1) and (2), the resonant absorption follows $A = 2Q_{leak}^{-1}Q_{loss}^{-1}/(Q_{leak}^{-1} + Q_{loss}^{-1})^2$, which predicts a maximum absorption coefficient. The excellent agreement between this theoretical prediction and the simulated spectra firmly validates the accuracy of the TCMT method. Figure 3(d) displays the simulated normalized surface impedance of the dark-mode system, with the black and red curves denoting the real and imaginary components. At the 365 Hz resonance, the impedance evaluates to $0.4 + i0.003$. The vanishingly small imaginary part (0.003) indicates a purely resistive resonance, which mitigates reactance-induced phase mismatch and facilitates the penetration of incident waves into the cavity. Simultaneously, the pronounced real part (0.4) reflects a strong intrinsic dissipation capability, efficiently converting the confined acoustic energy into heat through thermoviscous losses. This specific impedance behavior, near-zero reactance coupled with high dissipation, verifies the dark-mode characteristics.

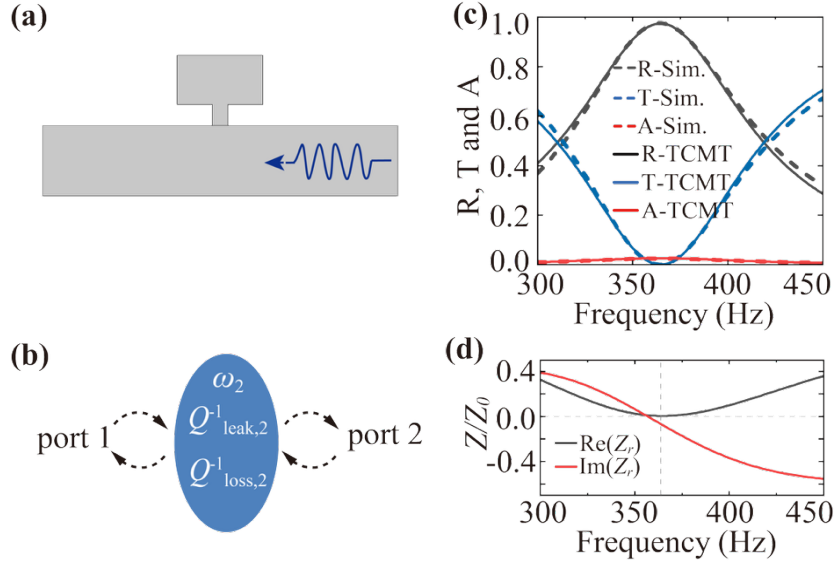

**Fig. 4. The bright mode resonator.** (a) Schematic of the waveguide solely coupled to the bright mode resonator (Helmholtz resonator). (b) Schematic of the coupled mode theory model. (c) Simulated and theoretical reflection (black curve), transmission (blue curve), and absorption spectra (red curve). (d) Simulated normalized impedance.

Fig. 4(a) presents the schematic of the main waveguide coupled solely to the bright-mode structure. This bright mode is realized by a conventional Helmholtz resonator (HR) comprising a short neck and a backing cavity, with its corresponding TCMT model illustrated in Fig. 4(b). As detailed in Fig. A.2 of Appendix A, the quality factors exhibit distinct dependencies on the geometric parameters of the Helmholtz resonator. Because the intrinsic loss is dominated by frictional acoustic dissipation inside the HR neck, widening the neck ($s_2$) significantly reduces $Q_{loss,2}^{-1}$. Conversely, since the leakage is dictated by radiative coupling through the neck into the main waveguide, an increase in $s_2$ provides a broader opening, thereby markedly enhancing $Q_{leak,2}^{-1}$ (Fig. A.2(b)). Meanwhile, adjusting the cavity width ($s_1$) has a relatively minor effect, with both loss and leakage showing only a slight decrease as $s_1$ increases (Fig. A.2(a)). By conducting parameter fitting at the target resonance frequency of $f_2 = 365$Hz, we ultimately extracted and finalized the geometric configuration to achieve the high-reflection objective, precisely matching the requisite quality factors: $Q_{leak,2}^{-1} = 0.298$ and $Q_{loss,2}^{-1} = 0.004$.

As depicted in Fig. 4(c), the reflection coefficient (black curve) approaches unity at the resonance frequency of $f_2 = 365\text{Hz}$. This near-perfect reflection is analytically demonstrated by Eq. (1), yielding an amplitude reflection coefficient of $r = -Q_{\text{leak}}^{-1}/(Q_{\text{leak}}^{-1} + Q_{\text{loss}}^{-1}) \approx -1$. This, coupled with the near-zero absorption (red curve), perfectly manifests the typical characteristics of a bright mode. The excellent agreement between the TCMT predictions (solid curves) and the FEM simulations (dashed curves) further validates the theoretical framework. For a deeper physical insight, Fig. 4(d) reveals an extremely low normalized surface impedance of $Z_r = 0.0008 - i0.077$ at resonance. The near-zero real part (black curve) at resonance implies that the system exhibits almost no intrinsic dissipation. Similarly, the imaginary part (red curve) is vanishingly small. This profound impedance mismatch induces high reflection, governed by $r = (Z_r - 1)/(Z_r + 1) \approx -1$. Consequently, this near-zero complex impedance establishes a perfect acoustically soft boundary, allowing the bright mode to effectively serve its intended purpose of blocking wave transmission.

## 2.3 Realization of asymmetric absorption

Having thoroughly investigated the acoustic responses of the waveguide coupled individually to either the dark- or bright-mode resonator, we now shift our focus to the integrated two-resonators system. When these two resonators, possessing distinct acoustic properties, are simultaneously introduced into the waveguide, the incident acoustic waves undergo not only fundamental transmission and reflection but also complex multiple scattering and near-field coupling effects between the resonators. To analyze the resulting asymmetric absorption mechanism, we utilize the transfer matrix method (TMM) [44]. The TMM effectively cascades the local acoustic properties of individual components through matrix multiplication to yield the macroscopic scattering coefficients of the overall system. For the coupled configuration in Fig. 1(b), the TMM relates the acoustic pressure and normal particle velocity across the system as follows:

$$\begin{pmatrix} p_{in} \\ -\boldsymbol{n} \cdot \boldsymbol{v}_{in} \end{pmatrix} = T_{\text{total}} \begin{pmatrix} p_{out} \\ -\boldsymbol{n} \cdot \boldsymbol{v}_{out} \end{pmatrix}, \tag{3}$$

where $p_{\text{in}}$ and $p_{\text{out}}$ denote the acoustic pressures at the inlet and outlet boundaries of the coupled system, respectively; $\boldsymbol{v}_{\text{in}}$ and $\boldsymbol{v}_{\text{out}}$ represent the corresponding particle velocity vectors; and $\boldsymbol{n}$ is the unit normal vector at the interfaces. The total transfer matrix $T_{\text{total}}$ can be derived by cascading the transfer matrices of the individual sub-components in sequence:

$$T_{\text{total}} = \begin{pmatrix} T_{11} & T_{12} \\ T_{21} & T_{22} \end{pmatrix} = T_{\text{resonator,1}} T_{\text{tube}} T_{\text{resonator,2}}. \tag{4}$$

Initially, utilizing the single-resonator reflection and transmission coefficients $r_i$ and $t_i$ ($i=1,2$) from Eqs. (1) and (2), the local acoustic response of the $i$-th resonator at the boundary is cast into a transfer matrix $T_{\text{resonator},i}(i = 1,2)$ [15,26]

$$T_{\text{resonator},i} = \begin{pmatrix} \frac{1-r_i^2+t_i^2}{2t_i} & \frac{[(1+r_i)^2-t_i^2]Z_0}{2t_i} \\ \frac{(1-r_i)^2-t_i^2}{2t_iZ_0} & \frac{1-r_i^2+t_i^2}{2t_i} \end{pmatrix}, \tag{5}$$

where $Z_0 = \rho_0 c_0$ denotes the acoustic impedance of air. Thus, the acoustic behavior of the coupled system in Fig. 1(b) is effectively described by the TMM alongside the leakage and loss factors. The matrix $T_{\text{tube}}$ dictates the phase accumulation during

wave propagation along the waveguide section separating the two resonators, given by:

$$T_{\text{tube}} = \begin{pmatrix} \cos(k_0 d) & iZ_0 \sin(k_0 d) \\ i\sin(k_0 d)/Z_0 & \cos(k_0 d) \end{pmatrix}, \quad (6)$$

with $k_0$ being the wavenumber in air and the physical distance $d$ between the resonators. By evaluating the global transfer matrix $T_{\text{total}}$, the intricate internal coupling and multiple scattering effects are implicitly captured within the matrix elements. Through straightforward algebraic manipulations, we can extract the complex transmission and reflection coefficients for acoustic waves incident from different directions as follows:

$$r_{\text{left}} = \frac{T_{11} + T_{12}/Z_0 - T_{21}Z_0 - T_{22}}{T_{11} + T_{12}/Z_0 + T_{21}Z_0 + T_{22}}, \quad (7)$$

$$r_{\text{right}} = \frac{-T_{11} + T_{12}/Z_0 - T_{21}Z_0 - T_{22}}{T_{11} + T_{12}/Z_0 + T_{21}Z_0 + T_{22}}, \quad (8)$$

$$t = \frac{2}{T_{11} + T_{12}/Z_0 + T_{21}Z_0 + T_{22}}. \quad (9)$$

Here, $r_{\text{left}}$ ($r_{\text{right}}$) denotes the reflection coefficient for acoustic waves incident from the left (right) port. Because the system's spatial configuration comprises two distinct resonators, its spatial inversion symmetry (mirror symmetry) is inherently broken. This symmetry breaking dictates that the main diagonal elements of the global transfer matrix $T_{\text{total}}$ are generally unequal (i.e., $T_{11} \neq T_{22}$). Consequently, acoustic waves incident from the left and right sides encounters entirely different impedance matching conditions, which naturally yields disparate reflection coefficients ($r_{\text{left}} \neq r_{\text{right}}$). Finally, governed by the principle of energy conservation, the absorption coefficient for left- or right-incident acoustic waves can be evaluated using the following equation:

$$A = 1 - |r_{\text{left}}(r_{\text{right}})|^2 - |t|^2. \quad (10)$$

Figure 5(a) presents the acoustic performance when sound waves are incident from the left port. In this scenario, the system exhibits near-zero reflection (black curve) and transmission (blue curve) at 365 Hz, thereby inducing near-perfect absorption (red curve) that approaches 99%. However, for incidence from the right port, as depicted in Fig. 5(c), the system maintains an identical transmittance (blue curve) dictated by the reciprocity principle, while the incident waves are predominantly reflected, reaching a reflectance of approximately 98% (black curve). The underlying asymmetric absorption mechanism can be elucidated as follows: upon left incidence, the reflected waves originating from resonator 1 and resonator 2 undergo destructive interference, resulting in the near-zero overall reflection. Conversely, sound waves incident from the right port are directly blocked by resonator 2 (the bright mode). Therefore, this pronounced asymmetric absorption is fundamentally governed by the direction-dependent reflectance. These results validate the proposed structure as an excellent asymmetric absorber/reflector. Furthermore, the theoretical TCMT predictions (solid curves) and the numerical FEM simulations (dashed curves) demonstrate basic agreement.

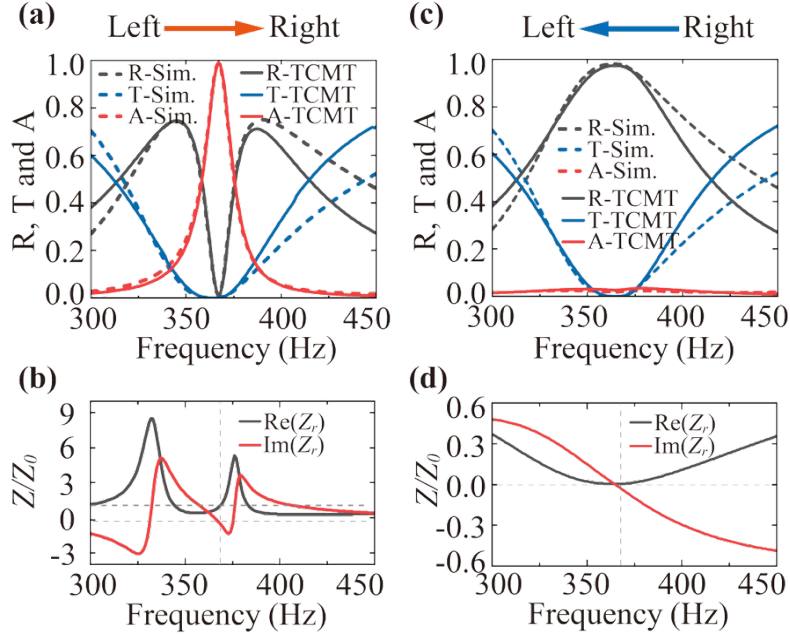

**Fig. 5. Asymmetric acoustic absorption performance.** Reflection (black curves), transmission (blue curves), and absorption (red curves) coefficient of the asymmetric system when sound waves incident from (a) the left port or (c) the right port. The solid and dashed curves represent the theoretical results based on TCMT and the numerical simulation results, respectively. (b) and (d) The respective normalized impedances corresponding to (a) and (c).

We simulated the system's normalized input impedance to corroborate these findings. As shown in Fig. 5(b), with the black and red curves denoting the real and imaginary components, left-incident waves encounter an impedance nearly perfectly matched to the air medium at the 365 Hz resonance, facilitating perfect absorption. Conversely, right incidence [Fig. 5(d)] yields a near-zero surface impedance at the same frequency. This extreme impedance mismatch effectively acts as an acoustically soft boundary, thereby directly inducing the observed near-total reflection.

To visually elucidate the asymmetric absorption, we present the simulated sound pressure, power dissipation density (PDD), and particle velocity fields in Fig. 6. Under left incidence [Fig. 6(a)], both the SCR and the Helmholtz resonator (HR) are excited, leading to strong acoustic energy localization (top panel). However, as evidenced by the PDD profile (middle panel), the energy is predominantly dissipated within the SCR, directly corroborating the dark mode's primary role in governing energy consumption. Furthermore, the bottom panel of Fig. 6(a) plots the distributions of the absolute sound pressure $|p|$ and particle velocity $v$ along the waveguide's centerline. It is well established that acoustic waves impinging upon a soft boundary naturally form a pressure node and a velocity antinode at the interface. The $|p|$ profile (solid black curve) clearly exhibits a node precisely at the position of resonator 2, accompanied by an antinode in the $v$ profile (solid red curve). This provides compelling physical evidence that the bright mode, represented by the HR, effectively functions as an acoustically soft boundary. Conversely, the pronounced standing wave pattern observed in Fig. 6(b) under right-incidence confirms that the SCR remains almost entirely unexcited. While a minor fraction of the acoustic energy is trapped by resonator 2, the corresponding dissipation is negligible (middle panel), resulting in the near-total reflection of the incident waves. The corresponding centerline distributions (bottom panel) further verify that the bright mode once again acts as an equivalent soft boundary, inducing the strong reflection.

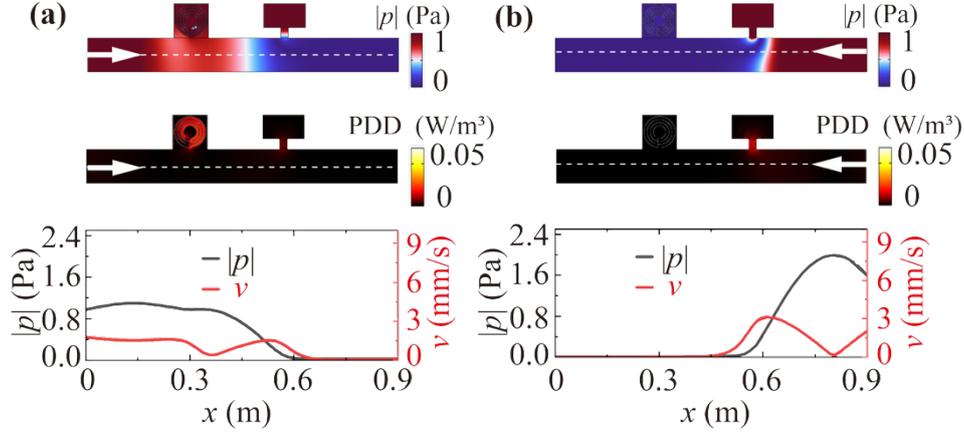

**Fig. 6. Field distributions at resonance.** Distribution of normalized sound pressure magnitude $|p|$ (top panel), total power dissipation density (PDD) (middle panel), and sound pressure (black curve) and particle velocity (red curve) on center line (bottom panel) at 365Hz when acoustic wave incident from (a) left port and (b) right port.

## 2.4 Tunable asymmetric absorption

Figure 7(a) illustrates the acoustic behavior of the acoustic waveguide coupled with a dark-mode resonator, revealing the variation of the leakage factor $Q_{\text{leak},1}^{-1}$ (black dotted line) and $Q_{\text{loss},1}^{-1}$ loss factor (red dotted line) as a function of the rotation angle $\theta$ of the SCR. Notably, $Q_{\text{loss},1}^{-1}$ remains strictly invariant across all rotation angles. This constancy arises because the intrinsic acoustic attenuation predominantly stems from thermoviscous dissipation within the narrow coiling channels. Since the rotation alters only the spatial orientation of the SCR without modifying its internal geometric topology, effective channel length, or cross-sectional area, the loss factor is entirely independent of $\theta$. Conversely, $Q_{\text{leak},1}^{-1}$ characterizes the radiation coupling strength between the dark mode resonator and the waveguide continuum, which is highly sensitive to the local acoustic microenvironment near the aperture. Consequently, $Q_{\text{leak},1}^{-1}$ exhibits a strong angle dependence, reaching its maximum at $\theta = 180°$ and its minimum at $\theta = 0°$.

This angle dependence directly affects the macroscopic acoustic absorption of the system. Figure 7(b) plots the acoustic absorption spectra of the SCR at various rotation angles. The theoretical predictions derived from the TCMT (solid curves) exhibit excellent agreement with the simulation result (dashed curves). For the 0°, 90°, and 180° rotational states (represented by black, blue, and red curves, respectively), a striking phenomenon emerges: while the peak absorption magnitude consistently remains clamped at approximately 0.5, the corresponding resonance frequency undergoes a pronounced redshift, migrating from 365 Hz down to 327 Hz. This feature demonstrates that a simple mechanical rotation can effectively shift the operating frequency band, offering a compelling paradigm for designing continuously tunable acoustic devices.

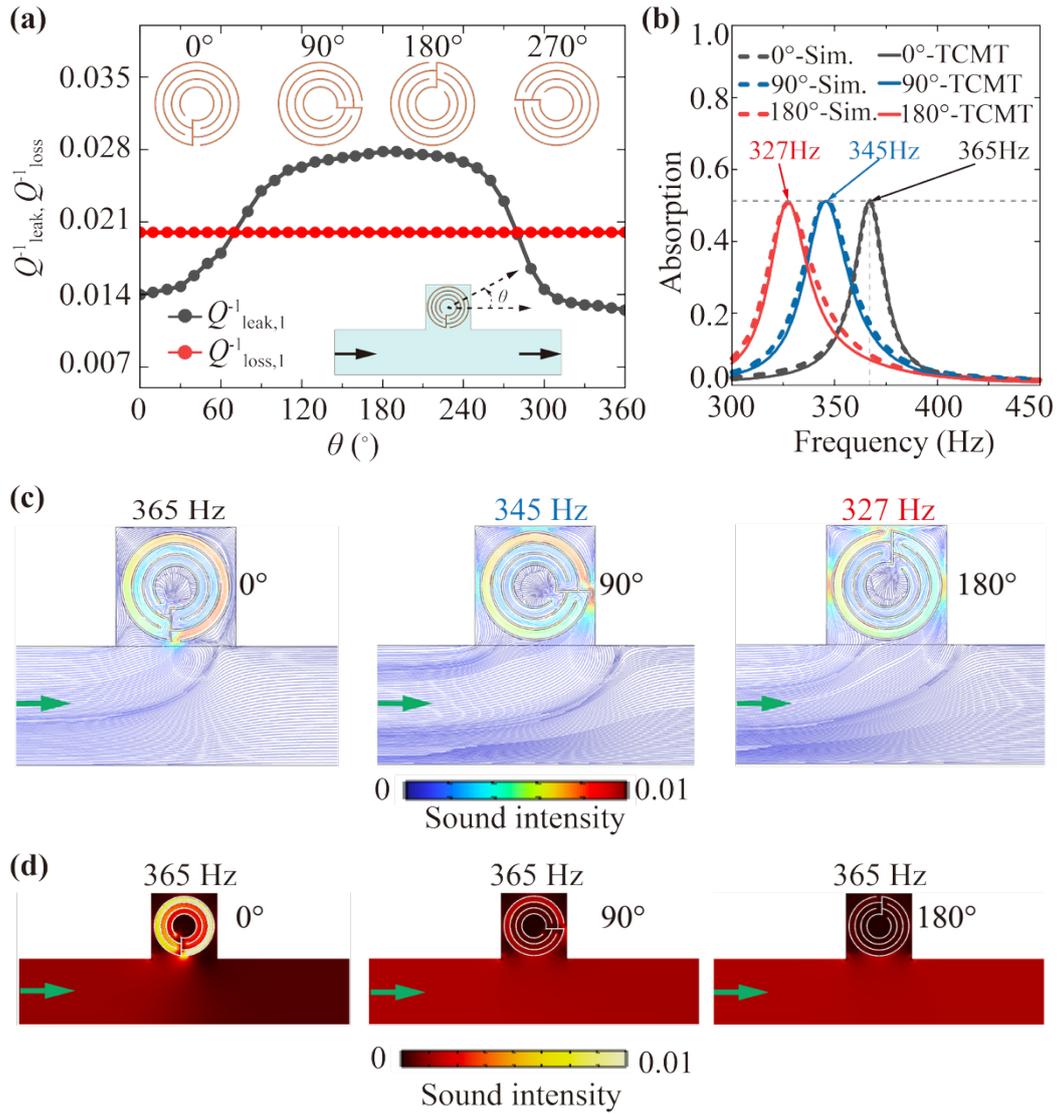

**Fig. 7. Angular tunability of the dark-mode acoustic behavior.** (a) Evolution of the leakage factor ($Q_{\text{leak},1}^{-1}$, black dotted curve) and loss factor ($Q_{\text{loss},1}^{-1}$, red dotted curve) as a function of the counterclockwise rotation angle $\theta$. Insets illustrate the aperture orientations at specific angles (top) and the geometric definition of $\theta$ (bottom right). (b) Simulated (dashed curves) and TCMT (solid curves) absorption spectra at different $\theta$, displaying a pronounced resonance redshift. (c) Local sound intensity streamlines at respective resonance frequencies (365, 345, and 327 Hz), revealing the elongation of the effective acoustic path with increasing $\theta$. (d) Sound intensity distributions at different rotation angles for a fixed operating frequency of 365 Hz.

To study the physical mechanism underlying this resonance shift, Fig. 7(c) illustrates the local sound intensity streamlines at the respective peak frequencies (365, 345, and 327 Hz). At 0° (left panel), the aperture of the SCR directly faces the main waveguide, allowing the acoustic streamlines to enter perpendicularly and smoothly. This straightforward coupling minimizes the physical propagation path, yielding resonance at a higher frequency (365 Hz). As the SCR rotates, its aperture gradually turns away from the incident field. For the 90° (middle) and 180° (right) configurations, the streamlines are forced to undergo severe bending or even bypass the entire external contour of the SCR to converge into the coiling channel. Physically, this severe distortion of the external local flow field is equivalent to significantly elongating the effective acoustic path at the aperture, thereby increasing the added acoustic mass. According to resonance theory, such an elongation inevitably drives the resonant frequency toward lower regimes, perfectly corroborating the redshift observed in Fig. 7(b).

Additionally, the system exhibits an interesting dynamic response at a fixed frequency: rotating the SCR from 0° to 180° at 365 Hz induces a sharp drop in absorption from 0.5 to nearly zero. This capability of giant amplitude modulation at a single frequency exhibits tremendous potential for acoustic switch applications. Figure 7(d) unveils the underlying physics of this switching mechanism by mapping the sound intensity magnitude ($|I|$) at a fixed 365 Hz excitation. At $\theta = 0°$ (left panel), the 365 Hz excitation perfectly matches the intrinsic resonance frequency of the system, leading to strong excitation of the SCR. Consequently, substantial acoustic energy is funneled into and highly localized within the coiling channel (indicated by the bright regions), where it is efficiently dissipated via thermoviscous effects, macroscopically manifesting as high absorption. However, when the SCR is rotated to 90° (middle panel) and 180° (right panel), the inherent redshift of the resonance frequency places the system in a state of severe detuning under the 365 Hz incident wave. This detuning fundamentally suppresses the dark mode excitation. Incident waves fail to enter the SCR and instead transmit directly through the main waveguide. Thus, simple mechanical rotation effectively switching the system from a highly absorbing state to a highly transmitting one.

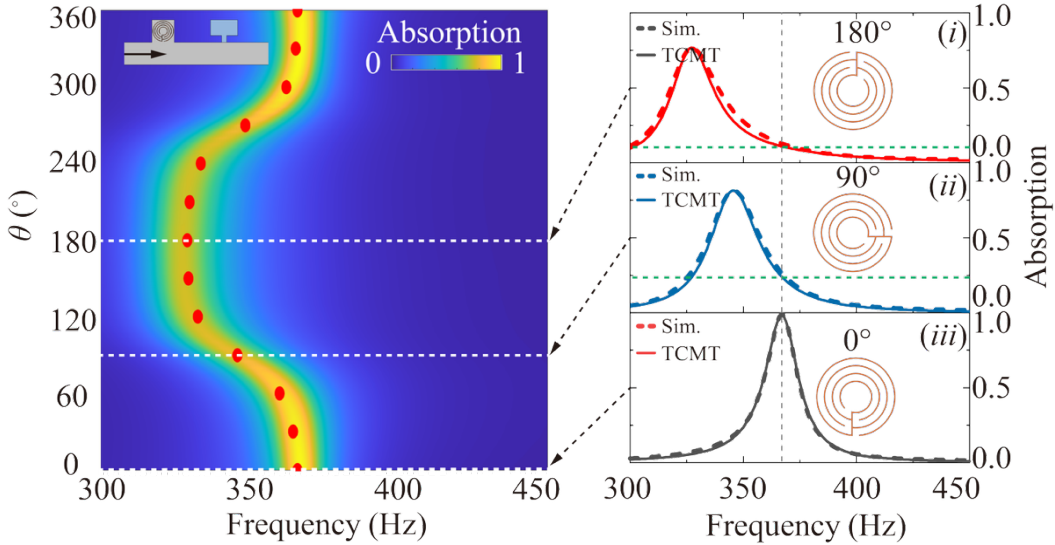

**Fig. 8. Tunable sound absorption of two resonator system.** The absorption coefficient versus frequency and SCR rotation angle $\theta$. Red dots mark the absorption peak positions predicted by the TCMT. The top-left inset schematically shows the geometric configuration of the two-resonator system, with the black arrow indicating the incident wave direction. (Right) Absorption spectra for $\theta = (i) 180°$, $(ii) 90°$ and $(iii) 0°$. Solid and dashed curves denote the TCMT calculations and FEM simulations, respectively. Subpanel insets display the corresponding SCR orientations.

Next, we extend our investigation to a comprehensive system comprising two coupled resonators. In this system, the bright mode exhibits a relatively high leakage factor, enabling its resonance response to cover a broad frequency band (as shown in Fig. 4). Consequently, within the target frequency range of interest, the acoustic properties of the bright mode remain essentially invariant, thereby continuously providing a stable acoustic soft boundary for the main waveguide [45]. Figure 8 illustrates the continuous dynamic evolution of the total absorption spectrum as a function of the SCR rotation angle $\theta$ within the dark mode (the top-left inset depicts the geometric configuration for left-incidence). The red dots trace the theoretical absorption peak positions calculated via TCMT, using the loss and leakage factor extracted from Fig. 7(a). To distinctly describe the physical responses at specific configurations, the right panels ($i$)-($iii$) of Fig. 8 plot the acoustic absorption spectra at $\theta$ =180°, 90° and 0°, respectively. Notably, at a fixed operating frequency of 365 Hz (marked by the vertical gray dashed line), the system achieves near-perfect absorption approaching 1.0 in the 0° rotation (Fig.8-$iii$). However, upon rotating the SCR to 90°, the absorption coefficient at this target frequency abruptly drops to 0.23 (indicated by the green horizontal dashed line, Fig. 8-$ii$). Further rotation to 180° causes the absorption to plummet to merely 0.08 (Fig. 8-$i$). This simple mechanical rotation suppresses the single-frequency energy absorption by ~92% (equivalent to nearly 11 dB of sound pressure level attenuation). Such a dynamic performance perfectly validates our proposed physical concept of a tunable acoustic switch enabled by rotational manipulation.

Furthermore, the total absorption peak of the two-resonator system exhibits remarkable dynamic tunability. Specifically, the peak frequency undergoes a pronounced redshift from 365 Hz at $\theta = 0°$ to 345 Hz at 90°, and further to 327 Hz at 180°. This trend perfectly mirrors the isolated dark mode behavior (Fig. 7), conclusively proving that the global frequency shift is directly governed by the SCR's local flow field distortion. In conventional designs [15,20,22,26,27,33,35,45], tuning the operating frequency typically necessitates altering the physical dimensions or structural parameters of the device. Here, however, a wide-range reconfiguration of the operating frequency is achieved solely by manipulating the spatial orientation of a single resonating unit. This paradigm offers fresh theoretical insights for designing deformation-free, continuously tunable intelligent acoustic metasurfaces.

## 2.5 Broadband absorption

In conventional broadband acoustic designs [15,20,22,26,27,33,35,44] it is typically necessary to parallel-couple multiple resonators with graded geometric dimensions to cover adjacent resonance frequencies. Such gradient designs inevitably increase the physical volume and manufacturing complexity of the system. To overcome this limitation, we propose a different broadband absorption strategy based on an array of isomorphic dark modes. As illustrated in Fig. 9(a), four SCRs with completely identical internal geometries are parallel coupled to the upper and lower boundaries of the main waveguide. Together with a downstream bright mode, they construct the broadband sound-absorbing system, where the inter-cell spacing along the $x$ direction is optimized to coordinate near-field coupling. The enlarged view in the red dashed box reveals the core mechanism of this design: the four isomorphic SCRs are assigned distinct spatial rotation angles (0°, 95°, 80°, and 180°, respectively).

Figure 9(b) illustrates the absorption performance of this multi-resonator system. The color-coded dashed lines depict the discrete absorption peaks contributed by the individual SCRs at their specific rotation angles. As governed by the physical mechanism validated in Figs. 7 and 8, the varying rotation angles introduce different degrees of effective acoustic path elongation. This effect finely separates the resonance frequencies of the four geometrically identical modules. More importantly, through rational spatial arrangement, these discrete resonance peaks undergo efficient superposition and synergistic merging. The thick red solid line in Fig. 9(b) plots the total absorption spectrum of the synergistically coupled system, clearly revealing the formation of a continuous, high-efficiency absorption band (highlighted by the green shaded region). The system achieves a quasi-perfect absorption ($A \geq 0.8$) from 325 Hz to 375 Hz, yielding an absolute bandwidth of 50 Hz and a fractional bandwidth of ~14.3%. By simply applying different rotation angles to identical SCRs, we achieve customized spectral merging for broadband absorption. This highlights the tremendous potential of these tunable dark modes in designing modular, fabrication-friendly acoustic metasurfaces.

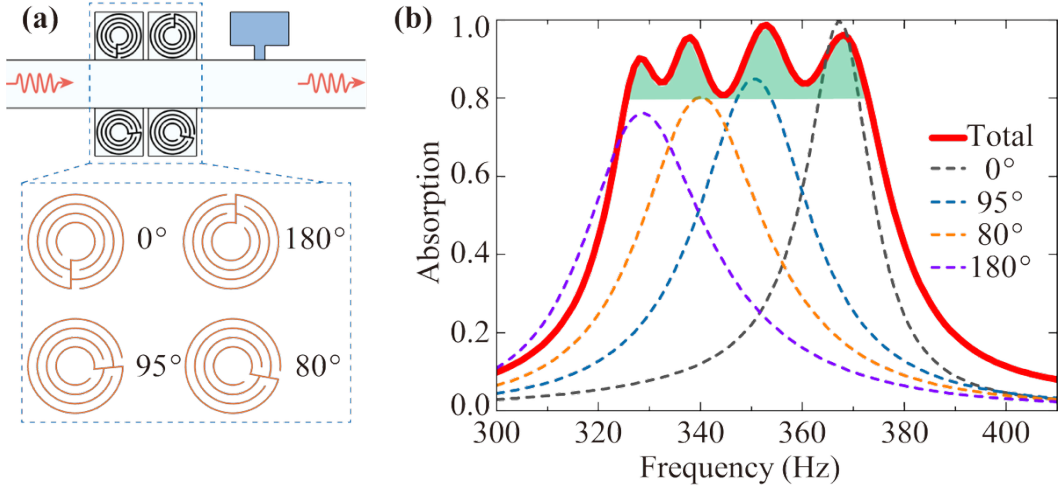

**Fig. 9. Broadband absorption based on a tunable ventilated metasurface.** (a) Schematic of the broadband metasurface, comprising four parallel SCRs (dark modes) and a downstream conventional resonator (bright mode). Inset: four identical SCRs arranged with distinct rotation angles (0°, 95°, 80°, 180°). (b) Broadband acoustic absorption spectra of the system. Color-coded dashed curves represent the discrete absorption peaks of the individual SCRs. The thick red solid line denotes the total absorption of the synergistic system. The green shaded region highlights the high-efficiency broad bandwidth (absorption coefficient $\geq 0.8$).

## III. CONCLUSION

In summary, we demonstrate a tunable low-frequency asymmetric ventilated absorber enabled by coupling a space-coiling resonator (SCR) with a Helmholtz resonator. By using the strong thermoviscous dissipation within the narrow-folded channels of the SCR in the dark mode, combined with the extremely low-impedance equivalent soft boundary characteristics of the bright mode, the system achieves a highly asymmetric, near-perfect absorption and reflection response at 365 Hz. Theoretical modeling via temporal coupled-mode theory (TCMT) and the transfer matrix method (TMM), supported by numerical simulations, reveals that simply rotating the SCR dynamically tunes its radiation leakage factor. This induces a substantial resonant frequency redshift without requiring any structural deformation. Notably, at a fixed 365 Hz excitation, this rotational mechanism acts as an efficient acoustic switch, yielding a massive amplitude modulation of ~92% in sound absorption. Ultimately, parallel-coupling four isomorphic SCRs at distinct rotation angles yields a broadband asymmetric absorber spanning 325 Hz to 375 Hz with a 14.3% absorption bandwidth (absorption ≥ 0.8). This work broadens the physical paradigms of asymmetric wave manipulation, drastically cuts manufacturing costs, and provides a scalable blueprint for compact, dynamically reconfigurable noise-control devices.

## ACKNOWLEDGMENTS

KL acknowledges fruitful discussions related to this work with Rfaqat Ali, Lijuan Fan, He Liu, Fathi Almobi, Iman Madkhali, Long Sun and Mukhammad Smagulov. The work described here is supported by the Office of the Sponsored Research (OSR) at King Abdullah University of Science and Technology (KAUST) under grant No. ORFS-CRG11-2022-5055, ORFS-OFP-2023-5560, and BAS/01/1626-01-01.

## DATA AVAILABILITY

The data that support the findings of this article are not publicly available. The data are available from the authors upon reasonable request.

**Appendix A. Modulation of dark and bright mode $Q$-factors based on TCMT**

Figure A.1 details how the key geometric parameters of the dark-mode resonator—spiral wall thickness $t$ and channel width $w$—modulate the system's leakage ($Q_{\text{leak},1}^{-1}$, black curves) and loss ($Q_{\text{loss},1}^{-1}$, red curves) factors when solely coupled to the main waveguide [corresponding to Fig. 3(a)]. Increasing $t$ [Fig. A.1(a)] reduces $Q_{\text{loss},1}^{-1}$ while keeping $Q_{\text{leak},1}^{-1}$ nearly constant. This suggests that varying $t$ alters the effective fluid volume-to-surface-area ratio, thereby independently tuning the thermoviscous-dominated intrinsic dissipation without perturbing the external radiative coupling. Conversely, increasing $w$ [Fig. A.1(b)] also decreases $Q_{\text{loss},1}^{-1}$ but induces a slight rise in $Q_{\text{leak},1}^{-1}$, indicating that wider channels effectively mitigate thermoviscous losses in confined acoustic propagation. Utilizing these parametric trends, we achieve the target dark-mode coupling parameters of $Q_{\text{leak},1}^{-1} = 0.014$ and $Q_{\text{loss},1}^{-1} = 0.02$ by setting $t = 0.14$ cm and $w = 0.76$ cm (marked by green dashed lines).

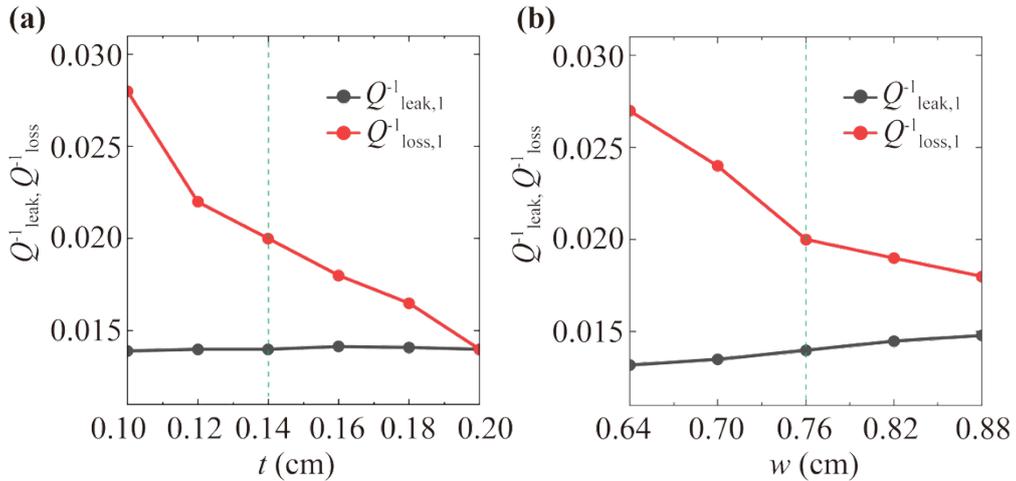

**Fig. A.1. Influence of structural parameters on Q factor.** Variations of $Q_{\text{leak},1}^{-1}$ (black curve) and $Q_{\text{loss},1}^{-1}$ (red curve) as a function of the (a) wall thickness $t$ and the (b) channel width $w$ of the space coiling resonator of the dark mode.

Figure A.2 shows the leakage ($Q_{\text{leak},2}^{-1}$, black curves) and loss ($Q_{\text{loss},2}^{-1}$, red curves) factors of the isolated bright mode as functions of the cavity width $s_1$ and neck width $s_2$. While varying $s_1$ [Fig. A.2(a)] yields only a marginal decrease in both factors, $s_2$ [Fig. A.2(b)] plays a main role in tailoring the acoustic response. Increasing $s_2$ significantly reduces $Q_{\text{loss},2}^{-1}$ while elevating $Q_{\text{leak},2}^{-1}$. Physically, the intrinsic loss stems primarily from thermoviscous dissipation within the confined neck; widening this cross-section effectively alleviates such frictional losses. Simultaneously, a larger $s_2$ introduces a broader physical aperture, which substantially strengthens the radiative coupling to the main waveguide and consequently increases the leakage factor. we finally get the target dark-mode coupling parameters of $Q_{\text{leak},2}^{-1} = 0.298$ and $Q_{\text{loss},2}^{-1} = 0.004$ by setting $s_1 = 12.1$ cm and $s_2 = 2.2$ cm.

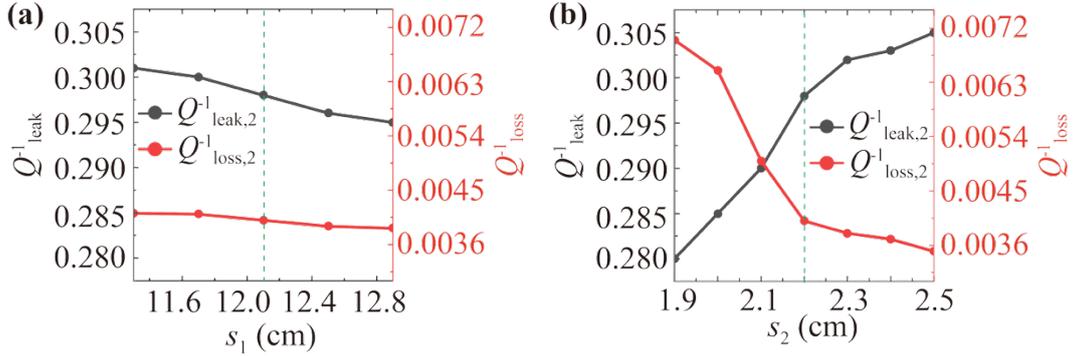

**Fig. A.2. Influence of structural parameters on Q factor.** Variations of $Q_{\text{leak},2}^{-1}$ (black curve) and $Q_{\text{loss},2}^{-1}$ (red curve) as a function of the (a) cavity width $s_1$ and the (b) neck width $s_2$ of the normal Helmholtz resonator of the bright mode.

**Appendix B. Theoretical impedance model for the two-resonator absorber via the transfer matrix method (TMM)**

**B.1. Equivalent impedance of the dark mode resonator $Z_{\text{dark}}$**

As detailed in the main text, the dark-mode resonator acting as a side branch to the main waveguide is a composite structure comprising a space-coiling resonator (SCR) and a square air cavity. Deriving its total equivalent impedance necessitates a sequential breakdown of its constituent acoustic responses. We begin by establishing the theoretical impedance model of the SCR.

B.1.1 Equivalent Helmholtz resonator (HR) model for the SCR

Acoustically, the complex space-coiling resonator (SCR) in Fig. B1(a) reduces to an equivalent low-frequency Helmholtz resonator (HR) model [Fig. B1(b)], where the neck (green region, bounded by rigid yellow walls) acts as a high-refractive-index medium. Physically, the internal folded channels significantly elongate the acoustic propagation path. This delayed transit to the central cavity macroscopically mimics a drastically reduced effective sound speed, thereby enabling efficient spatial folding and lowering the resonance frequency. The total unwrapped acoustic path length of this coiled channel (green line) is expressed as:

$$S_{\text{path}} = \sum_{0}^{N-1} 2\pi \cdot \frac{1}{S_{\text{seg}}} [r + (t + \frac{w}{2}) + (t + w)], \quad (\text{B.1})$$

where $N$ is the folding number, and $S_{seg}$ is the segment of SCR. Based on effective medium theory, the effective refractive index $n_r$ of this equivalent green neck is given by the ratio of the actual acoustic path to the straight radial geometric distance:

$$n_r = \frac{S_{path}}{R-r} = 20.75. \tag{B.2}$$

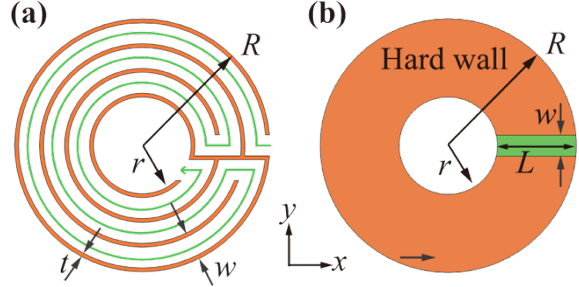

**Fig. B.1.** (a) Space-coiling resonator. (b) Equivalent Helmholtz resonator with a neck filled with a high-refractive-index medium.

According to classical acoustics, the total impedance of the equivalent Helmholtz resonator is the sum of its constituent components: $Z_{SCR} = Z_{neck} + Z_{cav}$. Here $Z_{neck} = i\omega M_{neck}$ and $Z_{cav} = 1/i\omega C_{cav}$ define the acoustic mass reactance of the neck and the acoustic compliance reactance of the cavity, respectively [46].

(a) Cavity impedance $Z_{cav}$

The central cavity acts as an acoustic compliance (analogous to a capacitor), modeling the elastic restoring force of the compressed air. Using $C_{cav} = V/\rho_0 C_0^2$, the cavity impedance is given by:

$$Z_{cav} = -i\frac{\rho_0 C_0^2}{\omega V}, \tag{B.3}$$

where $V = \pi r^2 H_z$ is the cavity volume, and the $z$ axial height $H_z$ is treated as a constant.

(b) neck impedance $Z_{neck}$

Given that the neck length $L \geq 0.1\lambda$, acoustic waves experience significant phase variations propagating through it. Consequently, the neck can no longer be simplified as a lumped-parameter mass element due to its pronounced frequency-dependent impedance. We therefore employ acoustic transmission line theory to re-derive the impedance model. The one-dimensional acoustic pressure $P(x)$ and volume velocity $U(x)$ within the neck are expressed as:

$$P(x) = Ae^{-ik_n x} + Be^{+ik_n x}, \tag{B.4}$$

$$U(x) = \frac{1}{Z_n}(Ae^{-ik_n x} - Be^{+ik_n x}), \tag{B.5}$$

where $A$ and $B$ denote the amplitudes of the forward-incident and backward-reflected waves, respectively, and $k_n$ is the effective wavenumber. The characteristic impedance is defined as $Z_n = \rho_n c_n/S$, with $S$ being the cross-sectional area and $\rho_n = \rho_0 \cdot n_r$ the effective density. To account for the non-negligible thermoviscous dissipation within the narrow channel, a complex sound speed is introduced: $c_n = c_0/n_r \cdot (1 + i\eta)$, where the loss factor is set as $\eta = 0.013$.

At the terminal ($x = L$), waves radiate from the small opening into a quasi-infinite external domain. This imposes a pressure-release boundary condition ($P \approx 0$, analogous to an acoustic short-circuit). Substituting $x = L$ yields:

$$P(L) = Ae^{-ik_nL} + Be^{+ik_nL} = 0 \tag{B.6}$$

which gives the amplitude relation $B = -Ae^{-2ik_nL}$. Evaluating the fields at the inlet ($x = 0$):

$$P(0) = A + B = A(1 - e^{-2ik_nL}), \tag{B.7}$$

$$U(0) = \frac{1}{Z_n}(A - B) = \frac{A}{Z_n}(1 + e^{-2ik_nL}). \tag{B.8}$$

The input impedance $Z_{neck} = P(0)/U(0)$, is finally derived as:

$$Z_{neck} = Z_n \frac{1 - e^{-2ik_nL}}{1 + e^{-2ik_nL}} = iZ_n \tan(k_nL) = i\frac{\rho_n c_n}{S}\tan(k_nL). \tag{B.9}$$

Notably, without the high-index medium ($n_r = 1$) in the low-frequency limit ($k_nL \ll 1$), the small-argument approximation $\tan(k_nL) \approx k_nL$ simplifies the neck impedance to $Z_{neck} \approx i\omega\frac{\rho_0 L}{S}$. Physically, the fluid column acts as a rigid oscillating mass, perfectly recovering the standard impedance of a classical Helmholtz resonator neck.

(c) Total impedance of SCR $Z_{SCR}$

To account for the additional acoustic mass induced by wave radiation at the aperture (i.e., end correction), the physical neck length must be appropriately modified. Moreover, the local flow field at the opening varies with the geometric rotation angle $\theta$, necessitating an angular correction. The effective neck length is thus defined as $L_{eff} = L + \delta_l + \delta_l(\theta)$ [46], comprising a constant correction $\delta_l = 0.065w$ and an angle-dependent term $\delta_l(\theta) = 0.0035\sin^2\frac{\theta}{2}$. Ultimately, the total SCR impedance becomes:

$$Z_{SCR} = Z_{neck} + Z_{cav} = i\frac{\rho_n c_n}{S}\tan(k_nL_{eff}) - i\frac{\rho_0 c_0^2}{\omega V}. \tag{B.10}$$

B. 1. 2. Combining SCR and square air cavity

As depicted in Fig. B.2(a), embedding the SCR into the side-branch square cavity creates a coupled dark-mode resonator. We aim to derive its input impedance, $Z_{dark}$, under isolated coupling with the main waveguide. The complex geometry [Fig. B.2(a)] simplifies to a waveguide model [Fig. B.2(b)], partitioning air region ③ into three sub-branches: branch a (the left slit between the SCR and cavity), branch b (the right curved air channel), and branch c (the top enclosed region).

Utilizing acoustic transmission line theory [47] and the electro-acoustic analogy, this system maps to an equivalent circuit network [Fig. B.2(c)]. Here, Region ① acts as a impedance $Z_1$. Region ② behaves as a variable cross-section waveguide; because its effective area depends on the SCR rotation angle $\theta$, it induces continuous impedance transformation along the wave path, resulting in an equivalent impedance $Z_2$. Electrically, the non-uniform branches a and b are in parallel, converging with the SCR impedance branch ($Z_{SCR}$) at the bottom (Node B) and with branch c at the top (Node C). Region c mimics a uniform waveguide terminated by a rigid wall, yielding impedance $Z_c$. Finally, the total surface impedance of the side branch is evaluated via the impedance translation method.

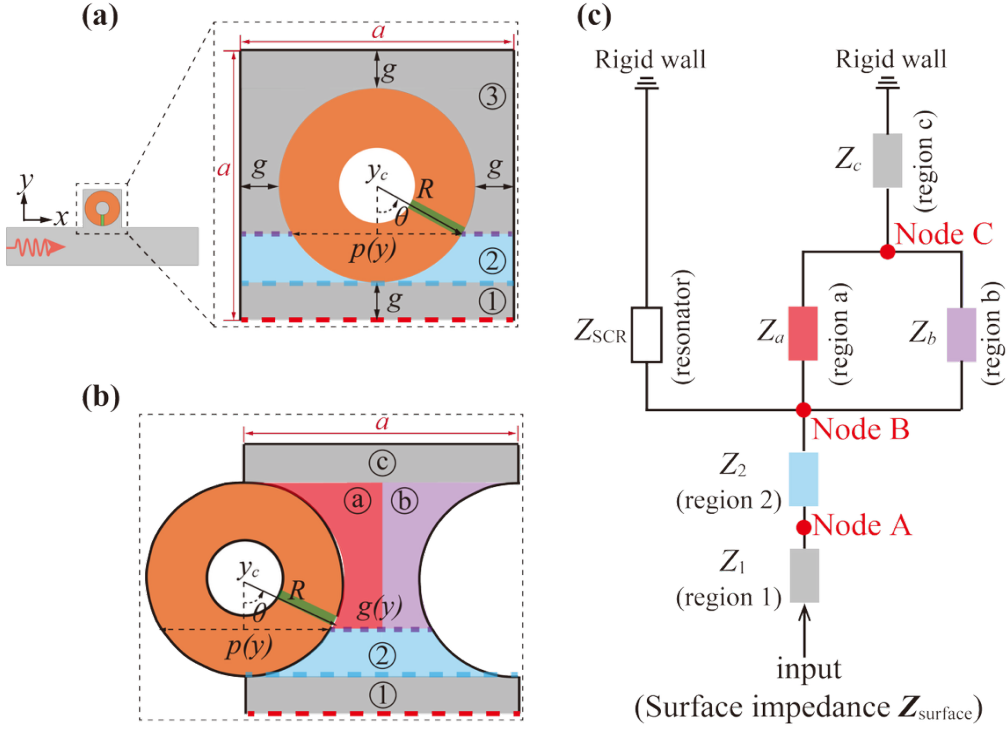

**Fig.B.2 Theoretical model of the rotation-tunable dark mode.** (a) Schematic geometry of the rotatable equivalent model configuration coupled to a main waveguide, where $\theta$ denotes the rotation angle. (b) Equivalent schematic of the partitioned physical regions. (c) The corresponding equivalent acoustic impedance circuit.

For the uniform waveguide in region c (with thickness $g$, cross-sectional area $S_{wg} = a \cdot H_z$, and characteristic impedance $Z_{wg} = \rho_0 c_0 / S_{wg}$), the transfer matrix is given by:

$$M_c = \begin{pmatrix} \cos(k_0 g) & iZ_{wg}\sin(k_0 g) \\ i\sin(k_0 g)/Z_{wg} & \cos(k_0 g) \end{pmatrix}. \quad (B.11)$$

In contrast, regions a and b (spanning from Node C to Node B) exhibit a $y$-dependent variable cross-section. To analytically model this, the domain is discretized along the $y$-axis into $N$ (e.g., $N = 50$) ultra-thin uniform slices, each with a thickness of $\Delta y$. The transfer matrix for the $j$-th slice is constructed as:

$$M_j = \begin{pmatrix} \cos(k_0 \Delta y) & iZ_{c,j}\sin(k_0 \Delta y) \\ i\sin(k_0 \Delta y)/Z_{c,j} & \cos(k_0 \Delta y) \end{pmatrix}, \quad (B.12)$$

where the local cross-sectional area is $S_j = (a - p(y)_j)H_z$, with $p(y)_j = 2\sqrt{R^2 - y_j^2(\theta)}$, yielding a local characteristic impedance $Z_{c,j} = \rho_0 c_0 / S_j$. By cascading these local matrices, the global transfer matrix $M_{ab}$ for regions a and b is obtained:

$$M_{ab} = \prod_{j=1}^{M_{\text{lower}}} M_j, \quad (B.13)$$

where $M_1$ and $M_{\text{lower}}$ denote the transfer matrices of the slices adjacent to Node C

and Node B, respectively. At the SCR junction, pressure continuity and velocity dictate a parallel electrical analogy, defining the matrix as $M_{\text{SCR}} = \begin{pmatrix} 1 & 0 \\ 1/Z_{SCR} & 1 \end{pmatrix}$.

For Region 2, another non-uniform waveguide, yields $M_2 = \prod_{j=1}^{M_{\text{lower}}} M_j$ via the identical stratification method (from Node B to Node A). Region 1, being uniform like region c, possesses the matrix $M_1 = \begin{pmatrix} \cos(k_0 g) & iZ_{wg}\sin(k_0 g) \\ i\sin(k_0 g)/Z_{wg} & \cos(k_0 g) \end{pmatrix}$. The global transmission relation for the dark mode is thus:

$$\begin{pmatrix} p_{in} \\ U_{in} \end{pmatrix} = M_{\text{total}} \begin{pmatrix} p_{end} \\ U_{end} \end{pmatrix}, \tag{B.14}$$

where $M_{\text{total}} = M_1 \cdot M_2 \cdot M_{\text{SCR}} \cdot M_{ab} \cdot M_c$. Given that the waveguide terminal (the top of region c) acts as a rigid load (i.e., $U_{\text{end}} = 0$), the ultimate surface impedance of the system simplifies to:

$$Z_{\text{surface}} = \frac{M_{\text{total},11}}{M_{\text{total},21}}. \tag{B.15}$$

This effectively maps the complex SCR-cavity side branch to a single parallel impedance $Z_{\text{surface}}$ loading the main waveguide. For the main waveguide impedance $Z_{\text{wg}} = \rho_0 c_0 / S_{wg}$, the total input impedance is $Z_{dark} = (Z_{\text{wg}} Z_{\text{surface}})/(Z_{\text{wg}} + Z_{\text{surface}})$. Substituting this into the fundamental reflection coefficient equation $r_{\text{dark}} = (Z_{dark} - Z_{\text{wg}})/(Z_{dark} + Z_{\text{wg}})$ yields:

$$r_{\text{dark}} = \frac{-Z_{\text{wg}}}{Z_{\text{wg}} + 2Z_{\text{surface}}}, \tag{B.16}$$

which simultaneously dictates the transmission coefficient $t_{\text{dark}} = 1 + r_{\text{dark}}$ as

$$t_{\text{dark}} = \frac{2Z_{\text{surface}}}{Z_{\text{wg}} + 2Z_{\text{surface}}}. \tag{B.17}$$

Subsequently, dictated by the principle of energy conservation, the absorption coefficient is calculated by:

$$A_{\text{dark}} = 1 - |r_{\text{dark}}|^2 - |t_{\text{dark}}|^2. \tag{B.18}$$

As depicted in Figs. B.3(c) and B.3(e), the theoretical spectra (solid lines) for both the acoustic coefficients ($R$, $T$, $A$) and the normalized impedance perfectly match the numerical simulations (dashed lines). This excellent agreement firmly validates the proposed analytical framework. Figure B.4 illustrates the rotational tunability of the dark mode absorption, where the resonance peak systematically shifts as the SCR rotates from 0° to 180°. Notably, the absorption spectra analytically derived from the theoretical impedance model (open circles) exhibit exceptional agreement with both the FEM SCR configuration (solid lines) and the equivalent medium model (dashed lines). This consistency firmly validates the reliability of the proposed impedance framework in capturing the rotational tuning behavior of the SCR.

**B2. Impedance of the bright mode resonator**

As depicted in Fig. B.3(b), the bright mode functions as a classical Helmholtz resonator and is analyzed via a lumped-parameter model [46]. Its acoustic mass is:

$$M_a = \frac{\rho_0 l_{eff,\text{bright}}}{S_b}, \tag{B.19}$$

with the neck cross-section $S_b = s_2 H_z$ and the effective neck length $l_{eff,\text{bright}} = 0.5s_2 + 0.78s_2$ (incorporating end corrections for flanged and unflanged terminations). The cavity acoustic compliance is:

$$C_a = \frac{V_b}{\rho_0 c_0^2}, \tag{B.20}$$

Where $V_b = s_2 h H_z$ is the cavity volume. Rapid air oscillation in the confined neck induces viscous frictional losses, introducing an acoustic resistance $R_{\text{viscous}} = l_{eff,\text{bright}}\sqrt{2\rho_0\omega\mu}/(S_b s_2)$ [48]. Thus, the total input impedance becomes:

$$Z_{\text{HR}} = R_{\text{viscous}} + i(\omega M_a - \frac{1}{\omega C_a}), \tag{B.21}$$

Acting as a parallel side-branch on the main waveguide, junction boundary conditions (pressure continuity and volume velocity conservation) dictate the reflection and transmission coefficients:

$$r_{\text{bright}} = \frac{-Z_{\text{wg}}}{Z_{\text{wg}} + 2Z_{\text{HR}}}, \tag{B.22}$$

$$t_{\text{bright}} = \frac{2Z_{\text{HR}}}{Z_{\text{wg}} + 2Z_{\text{HR}}}, \tag{B.23}$$

with the absorption coefficient $A_{\text{bright}} = 1 - |r_{\text{bright}}|^2 - |t_{\text{bright}}|^2$. As evidenced in Figs. B.3(d) and B.3(f), the analytical spectra (solid lines) perfectly match the numerical simulations (dashed lines), firmly validating the accuracy of the proposed impedance framework.

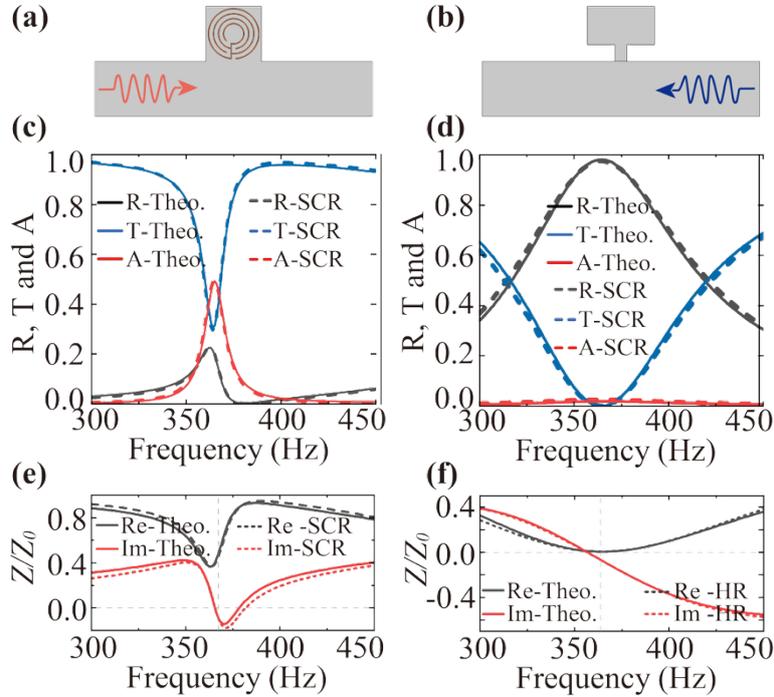

**Fig. B.3. Scattering and impedance characteristics of the dark and bright modes.** (a, c, e) Performance of the dark mode (SCR configuration) and (b, d, f) the bright mode (HR configuration) operating independently. Solid lines and dashed lines represent the theoretical impedance model and simulation results, respectively.

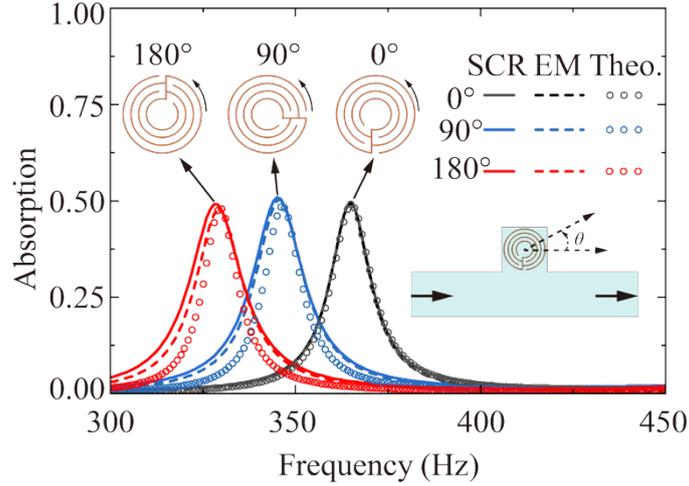

**Fig. B. 4. Dependence of dark mode absorption on rotation.** Absorption coefficient of the structural configuration (solid lines), the equivalent model (dashed lines), and the theoretical impedance model (open circles) at different rotation angles (0°, 90°, and 180°). Top insets depict the geometry of the SCR at corresponding angles; the middle inset illustrates the schematic of the resonator coupled to a waveguide.

**B.3. Absorption of the total system vis TMM**

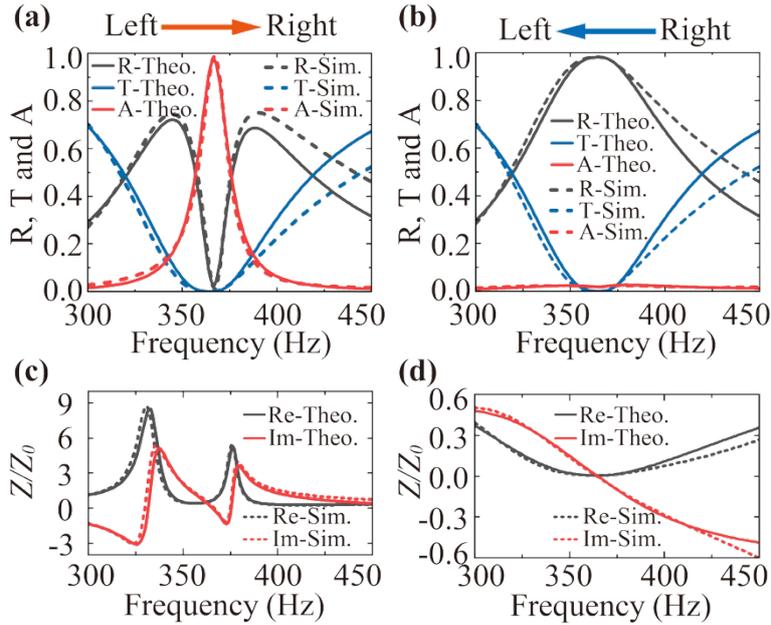

**Fig. B.5. Asymmetric acoustic absorption performance.** Reflection (black curves), transmission (blue curves), and absorption (red curves) coefficient of the asymmetric system when sound waves incident from (a) the left port or (c) the right port. The solid and dashed curves represent the theoretical impedance model results and the FEM results, respectively. (b) and (d) The respective normalized impedances corresponding to (a) and (c).

With the individual impedances ($Z_{\text{surface}}$ and $Z_{\text{HR}}$) derived, the two resonators act as shunt elements on the main waveguide. Imposing pressure continuity and velocity yields their point-transfer matrices: $T_{\text{dark}} = \begin{pmatrix} 1 & 0 \\ 1/Z_{\text{surface}} & 1 \end{pmatrix}$ and $T_{\text{bright}} =$

$\begin{pmatrix} 1 & 0 \\ 1/Z_{\text{HR}} & 1 \end{pmatrix}$. The intermediate straight tube segment of length $d$ is represented by $T_{\text{tube}} = \begin{pmatrix} \cos(k_0 d) & iZ_{wg}\sin(k_0 d) \\ i\sin(k_0 d)/Z_{wg} & \cos(k_0 d) \end{pmatrix}$. Cascading these matrices dictates the total system transfer matrix:

$$T_{\text{total}} = \begin{pmatrix} T_{11} & T_{12} \\ T_{21} & T_{22} \end{pmatrix} = T_{dark} T_{\text{tube}} T_{\text{bright}}, \tag{B.24}$$

Solving this concurrently with Eqs. (3) and (7)-(9) extracts the system's scattering and absorption coefficients in Fig. B. 5. The analytical predictions (solid lines) perfectly match the FEM results (dashed lines), firmly validating our theoretical model.